# Highly efficient rubrene-graphene charge transfer interfaces as phototransistors in the visible regime

*Gareth F. Jones [1], Rui M. Pinto [2], Adolfo De Sanctis[1], V. Karthik Nagareddy [1], C. David Wright [1], Helena Alves [3], Monica F. Craciun [1], Saverio Russo [1]\**

[1] Centre for Graphene Science, College of Engineering, Mathematics and Physical Sciences, University of Exeter, Exeter EX4 4QF, United Kingdom.
[2] INESC MN and IN, Rua Alves Redol No. 9, 1000-029 Lisboa, Portugal.
[3] CICECO - Aveiro Institute of Materials, Physics Department, University of Aveiro, 3810 Aveiro, Portugal

\* Correspondence: S. Russo, Email: S.Russo@exeter.ac.uk

**Atomically thin materials such as graphene are uniquely responsive to charge transfer from adjacent materials, making them ideal charge transport layers in phototransistor devices. Effective implementation of organic semiconductors as a photoactive layer would open up a multitude of applications in biomimetic circuitry and ultra-broadband imaging but polycrystalline and amorphous thin films have shown inferior performance compared to inorganic semiconductors. Here, we utilize the long-range order in rubrene single crystals to engineer organic semiconductor-graphene phototransistors surpassing previously reported photo-gating efficiencies by one order of magnitude. Phototransistors based upon these interfaces are spectrally selective to visible wavelengths and, through photoconductive gain mechanisms, achieve responsivity as large as $10^7 AW^{-1}$ and a detectivity of $9 \times 10^{11} \, Jones$ at room temperature. These findings point towards implementing low-cost, flexible materials for amplified imaging at ultra-low light levels.**

The planar interfaces formed between monolayer graphene and semiconductor materials present unique opportunities for amplified detection of weak light signals. In these systems, electron-hole pairs are excited in the semiconductor layer by an incident flux of photons ($\phi$) with energy equal to or greater than the optical band gap. Charge carriers of one polarity are then transferred from the semiconductor into graphene according to the electrochemical potential gradient at the interface, modulating the carrier concentration in graphene by $\Delta n$. Charge carriers of the opposite polarity remain in the semiconductor layer, often in localized trap states, and recombine after an average lifetime $\tau_L$. The associated probability encompassing the photo-excitation of charges and their transfer to graphene is the photo-gating quantum efficiency (PGQE), $\eta_{PG} = \Delta n/\phi\tau_L$. If an external voltage ($V_{DS}$) is applied along the length ($L$) of a graphene-semiconductor interface (**Figure 1**a), charge carriers transferred into graphene will drift between source and drain electrodes over an average time $\tau_{tr} = L^2/\mu V_{DS}$ where $\mu$ is the charge carrier mobility of graphene. For channels that are just a few microns long, trapping lifetimes are typically up to nine orders of magnitude longer than transit time-scales, resulting in a net photoconductive gain $G = \tau_L/\tau_{tr}$. The external quantum efficiency (EQE, $\eta_{EQE} = G\eta_{PG}$) of these phototransistors can therefore far exceed 100%, particularly if the PGQE is optimal, allowing electrical detection of femtowatt light signals at room temperature in micron-scale devices. Optimizing the EQE of graphene phototransistors through exploration of various material combinations is now a highly active field of research, with previous studies focusing on hybrid structures of graphene interfaced with colloidal quantum dots, [1-5] transition metal dichalcogenides, [6-8] III-VI semiconductors, [9] metal oxides, [10] perovskites, [11] chlorophyll, [12], organometallic complexes [13] and organic semiconductors thin films [14-17]. So far, room temperature EQE as large as $10^8$ electrons per photon has been reported in graphene/colloidal quantum dot phototransistors, with operational speeds suitable for video-rate imaging. [2].

Phototransistors that combine graphene with organic semiconductors are particularly desirable owing to the gamut of complementary properties found in these systems. The spectral selectivity of π-conjugated semiconductors can be tailored through chemical or structural modification to emulate the trichromatic response of cone cells in mammalian retina [18] or exhibit ultra-broadband UV-to-NIR sensitivity. [19] Additionally, organic semiconductors have an intrinsic affinity to biological systems that is vital for developing innovative healthcare sensors. However, organic semiconductor-graphene phototransistors [14-17] have shown radically inferior PGQE compared to inorganic semiconductor-graphene phototransistors, where $\eta_{PG}$~25% using colloidal PbS quantum dots [2]. Short exciton diffusion lengths (~10nm) [20] and inhomogeneity [21] present in organic semiconductor films are likely to play a critical role in limiting quantum efficiencies. Amorphous films of P3HT [16] have been shown to exhibit a $\eta_{PG}$~0.002%, whereas $\eta_{PG}$~0.6% has been achieved using polycrystalline films of epitaxially grown $C_8$-BTBT [17] (see Table S1), suggesting that the disorder of the crystal structure in graphene/organic semiconductor phototransistors plays a pivotal adverse role. Furthermore, the current operational speed of these hybrid devices is also far from ideal. Response times lasting many seconds make them too slow for imaging applications, yet not sufficiently stable to function as optical memories. Here, we address both of these challenges for the first time using a single crystal organic semiconductor, rubrene, as the light-absorbing layer in a graphene phototransistor. Long-range herringbone stacking of rubrene molecules in a single crystal facilitates exciton diffusion over several microns. [22, 23] We exploit this to achieve both efficient light absorption and efficient extraction of photo-excited charge carriers, resulting in an external and internal PGQE as high as 1% and 5% respectively. These organic single crystal-graphene phototransistors exhibit responsivity as large as $10^7 AW^{-1}$ at room temperature and a specific detectivity of $9 \times 10^{11} \, Jones$.

The device structure consists of a rubrene single crystal grown by physical vapor transport (see Methods) laminated onto a pre-fabricated graphene transistor (Figure 1b). After device fabrication, we used the well-established used method of cross-polarized optical microscopy [24] to demonstrate macroscopic molecular ordering and absence of polycrystalline domains across the rubrene single crystals. Figures 1c and 1d show uniform brightness across the entirety of the interface, with the magnitude of brightness dependent on the angle between the crystal's long axis and the polarization plane of incident light. Polar plots of the average brightness over three distinct interface regions (Figure 1e) revealed identical birefringence, which can only be present in a structurally pristine single crystal of rubrene. Figure 1f shows photoluminescence (PL) spectra from another rubrene crystal measured at two locations, one with and one without an underlying sheet of graphene. In both cases, the PL spectra fit well with two sets of equidistant Voigt functions representing the vibronic progression of radiative transitions polarized along the *L/N* ($< 2.05 eV$) and *M* ($> 2.05 eV$) axes of rubrene molecules (Figure 1g). [25] Although *M*-polarized emission is 10-20 times stronger, these peaks are suppressed in Figure 1f. This indicates that our axis of illumination/detection is oriented parallel to the *M* axes of rubrene molecules and, therefore, is normal to the *ab* crystal plane (Figure 1h). We confirm this crystallographic orientation via polarized Raman spectroscopy, shown in Figure S5 [27]. A PL band located at $1.91 eV$, which is a signature of photo-oxidation [25] and deep trap states [28], was absent from all measured samples confirming the high purity of these crystals. Comparing the two PL spectra, the presence of graphene underneath rubrene reduces the PL intensity by approximately 25%. This PL quenching suggests that a substantial fraction of excitons dissociate across the rubrene-graphene interface, although Förster resonance energy transfer could also cause PL quenching [29] and would be detrimental to the PGQE. In order to determine the efficiency with which electron-hole pairs dissociate at the rubrene-graphene interface, we proceeded to study the electrical response of a channel segment to flood illumination ($\lambda = 500 nm$).

Figure 1i shows that photocurrent ($I_{PH}$) measured from a rubrene-graphene channel segment has a linear dependence on source-drain voltage. This linearity is expected for systems exhibiting photoconductive gain $\left(G = \frac{\mu \tau_L V_{DS}}{L^2} \propto V_{DS}\right)$ and suggests that excitons in rubrene are dissociated at the interface with graphene. For a source-drain bias voltage of $30mV$ and a measured field-effect mobility of $1300 cm^2 V^{-1} s^{-1}$ (see Figure S8), we estimate the transit time to be $\tau_{tr} = 10ns$. In **Figure 2**a, we show the resistance of the same rubrene-graphene channel as a function of applied gate voltage under a variety of optical power densities ($P$). In all cases, charge transport along the interface is clearly dominated by the ambipolar behavior of monolayer graphene with illumination inducing an up-shift of the charge neutrality point ($\Delta V_{CNP}$). This up-shift is indicative of photo-excited holes being transferred from rubrene into graphene whilst electrons remain confined to the rubrene crystal. [2] A plot of the upward shift of the charge neutrality point as a function of power density in Figure 2b reveals that the photo-gating effect saturates for $P > 5Wm^{-2}$. Such an observation could originate from screening of the built-in field at the rubrene-graphene interface [2] or an increased probability of bimolecular recombination in rubrene [23] at high photo-excited charge carrier densities. To characterize the gain in these phototransistors, we focus on the transient response of this interface to weak light signals, where $P < 300 \mu W m^{-2}$, in the inset of Figure 2b. Whilst the rise time of the detector is relatively fast, taking approximately $100ms$ to reach steady-state conditions under illumination, the transition back to dark current levels lasts for tens of seconds and is indicative of the average lifetime of electrons localized in rubrene. [2] The photocurrent after illumination was fit with a bi-exponential decay function (pink dashed line), suggesting that at least two distinct lifetimes exist for electrons in rubrene [12]. Taking a weighted average of the two decay constants, we calculate an average lifetime of $\tau_L = 24 \pm 3$ seconds and gain of $G \approx 10^9$. The photoresponse of these devices is tunable with applied gate voltage and we observe a responsivity ($\gamma = I_{PH}LW/P$) as large as $1.4 \times 10^5 AW^{-1}$ and EQE= $3.4 \times 10^7$% under light levels equivalent to moonlit conditions.

In Figure 2e, we compare the spectral response of a rubrene-graphene interface with an equivalent sample solely comprised of rubrene. Identical voltages were applied to each device ($V_{DS} = 30mV$, $V_G = 0V$) and channel geometries were kept consistent ($L = 5\mu m$, $W = 90 - 100 \mu m$). Dashed lines show the simulated absorbance of the isolated crystal to be approximately twice that of the crystal in contact with graphene due to differences in thickness ($405 nm$ and $202 nm$ respectively). The two transistors produced responsivity spectra with very similar shapes, confirming that photocurrent signals in rubrene-graphene transistors arise purely from light absorption in the organic single crystal and that no new chemical species form at the interface. A comparison of the magnitude of responsivity in each device in the inset of Figure 2e shows that graphene/rubrene phototransistors exhibit values of $\gamma$ up to six orders of magnitude larger than in isolated rubrene. This observation demonstrates that the high charge carrier mobility in graphene plays an essential role for efficient transport of photo-excited holes between source and drain electrodes whereas the surface photoconductivity of rubrene [23] does not significantly contribute towards read-out signals.

Channel geometry is a significant, but extraneous, factor that is largely responsible for the large variations of responsivity amongst previously reported graphene-based phototransistors [2-17]. Indeed, a proportionality of $\gamma \propto G\eta_{PG} \propto L^{-2}$ is expected, given that $G \propto L^{-2}$ whilst $\eta_{PG}$ is independent of channel length. However, large lateral electric fields and possible exciton quenching effects at metallic source/drain electrodes could significantly affect the PGQE in shorter channels. To exclude these spurious effects, we have conducted for the first time a scaling experiment of $\gamma$ as a function of channel length in Figure 3a. We find that the responsivity shows the expected $L^{-2}$ dependence when normalized to the charge carrier mobility and potential difference ($V_{ch}$) across each segment in order to account for contact resistance and doping inhomogeneity (Figure S8). This demonstrates that the active area of the phototransistors comprises the whole rubrene-graphene interface between the source and drain electrodes. Hence, a more meaningful comparison of $\gamma$ for any graphene-based phototransistor can be achieved by accounting for the inverse square dependence on channel length, provided the PGQE is independent of *L*.

As previously mentioned, the PGQE of graphene-based phototransistors generally increases at lower absorbed photon densities. In Figure 3b, we explore the limit of this effect by measuring the non-linear power dependence of responsivity in a rubrene-graphene interface exposed to ultra-weak light signals. A maximum of $\gamma_{max} \approx 1 \times 10^7 AW^{-1}$ is reached for the lowest measured optical power densities. This marks the first report of responsivity comparable to the record room-temperature performance of inorganic semiconductor-graphene phototransistors [2, 6] from an entirely organic equivalent. Analogous to previous phototransistor studies, [2, 4, 6] we fit this non-linear power dependence with the function

$$\gamma = \frac{\gamma_{max}}{1+(P/P_0)^n} \quad (1)$$

where $P_0$ marks a threshold power below which the responsivity saturates, and $n$ is an exponent which dictates the decline in responsivity above this threshold. A best fit of Equation (1) (blue dashed line) yields $P_0 \approx 1.1 \mu W m^{-2}$ and $n = 0.70 \pm 0.04$. Re-arranging the expanded expression for responsivity, $\gamma = (e\eta_{PG} / h\nu)(\mu V_{DS}\tau_L / L^2)$, we are able to calculate the PGQE (blue data) shown on the right y-axis in Figure 3b. The simulated absorbance (*A*) of the rubrene crystal, shown in Figure 2e, is then used to calculate the internal photo-gating quantum efficiency (charges transferred to graphene per *absorbed* photon, orange data) as $\eta_{iPG} = \eta_{PG} / A$. For power densities equivalent to sub-femtowatt incident signals we calculate $\eta_{PG} \approx 1\%$ and $\eta_{iPG} \approx 5\%$. Comparatively, this value of PGQE is four orders of magnitude greater than a previous study that combined amorphous films of P3HT [16] with graphene and one order of magnitude higher than epitaxially grown polycrystalline films of $C_8$-BTBT grown on graphene [17]. We attribute the superior PGQE in rubrene-graphene interfaces to a combination of two factors. Firstly, a rubrene single crystal serves as an ideal light-absorbing layer due to the extremely low density of charge traps in the bulk of the crystal [30] and the large intermolecular overlap of pi orbitals which facilitates Dexter-type diffusion of triplet excitons over several microns. [22, 23] A far larger number of excitons are therefore able to diffuse to the graphene interface and dissociate. Secondly, this is the first study to examine the PGQE of organic semiconductor-graphene phototransistors at extremely low absorbed photon densities where bimolecular recombination and triplet-triplet fusion are not significant loss mechanisms. [23]

Previous studies of hybrid graphene phototransistors have attributed a decline in responsivity with increasing optical power to the saturation of available charge trap in the light-absorbing semiconductor layer. [4] Assuming this to be true in the case of rubrene-graphene interfaces, we use the threshold power density to estimate the density of trap states in rubrene available for photo-gating processes as $N_t = \eta_{iPG}\tau_L P_0/h\nu \approx 5 \times 10^8 cm^{-2}$. Even if graphene screens an overwhelming proportion of the traps otherwise present at the rubrene-SiO$_2$ interface ($\sim 10^{12} cm^{-2}$), [30] our estimate of $N_t$ is too low to be physically plausible. Hence, the density of available interface trap states does not govern the non-linear responsivity. Instead, we note that responsivity follows a power exponent of approximately $-2/3$ for $P \gg P_0$, which closely correlates with the signature of triplet-charge recombination from surface photoconductivity experiments on rubrene. [23] The onset of these interactions occurs at higher absorbed photon densities ($> 10^{15} cm^{-3} s^{-1}$) in isolated crystals but it is reasonable to expect a lower threshold considering the additional population of charge carriers from physical contact with graphene. This finding should help to inform future strategies of interface modification.

In Figure 3c, we operate a $5\mu m$ rubrene-graphene channel at a gate voltage of $V_G = 10V$ such that the Fermi level of graphene lies within the conduction band. With successive cycles of illumination, the drain current gradually drifts away from its original dark value due to the fall time of the detector exceeding the time under dark conditions. By applying a gate voltage pulse when the light source is extinguished, we momentarily reduce the built-in field across the interface which otherwise limits recombination of photo-excited electrons in rubrene [2] (Figure 3d). Using this technique, we surpass the bandwidth limitations that previous graphene-organic semiconductor phototransistors have suffered from resulting in a maximum detectivity of $9 \times 10^{11} Jones$ (see Supplementary Section S6).

A comparison of the responsivity measured in published state-of-the-art organic semiconductor-graphene phototransistors (**Figure 4**a) demonstrates that our devices attain record high values of responsivity at unprecedentedly low incident photon flux, comparable to that of inorganic equivalents (see Figure S11). However, since responsivity depends on extraneous parameters such as channel length, source-drain voltage and the charge carrier mobility of graphene, it is difficult to gain accurate insight as to the relative performance of each organic material solely from this figure of merit. Indeed, all of these extraneous parameters vary significantly amongst studies and influence photoconductive gain rather than the intrinsic photo-gating quantum efficiency of each material interface. Hence, we define $\eta_{EQE}L^2/\mu V_{DS} = \eta_{PG}\tau$ as a more appropriate figure of merit which is independent of $\mu$, $V_{DS}$ and $L$ (Figure 4b). This quantity reflects the maximum achievable EQE. Specific detectivity could also serve as an informative figure of merit but is often not reported and sometimes overestimated by assuming Shot-noise limited performance. The comparative plots of Figure 4 conclusively demonstrate that rubrene single crystal-graphene phototransistors are uniquely suited for amplified detection of extremely weak light signals in all-organic electronics, where the long-range diffusion of excitons in rubrene facilitates both high absorbance and efficient extraction of photo-generated charge carriers. Two caveats when considering possible applications of graphene-based phototransistors are the operational bandwidth and noise-equivalent power of each phototransistor. Whilst bandwidth can be improved through gate voltage modulation or screening of deep trap states, for example with ionic polymer gates, [8] it is the 1/f dark current noise that limits the noise-equivalent power of graphene detectors for bandwidths below $100 kHz$. [31] This limitation can be addressed to some degree by using single crystal graphene films [32] and one-dimensional electrode contacts. [33] Overall, the parallel efforts to develop optimal light-absorbing layers, improve operational bandwidth and reduce 1/f noise could enable graphene-based phototransistors to reach detectivity values rivalling those of single photon detectors.

In conclusion, interfaces of monolayer graphene and rubrene single crystals are promising systems for ultra-sensitive detection of visible light. Long-range order in rubrene crystals facilitates effective transfer of photo-generated charges to graphene with an external and internal efficiency of 1% and 5% respectively. Utilizing these interfaces as phototransistors, responsivity as high as $10^7 AW^{-1}$ can be achieved for sub-femtowatt incident signals, comparable to the record performance of graphene-quantum dot detectors. Finally, we emphasize the importance of distinguishing between the contributions of internal gain, photo-gating quantum efficiency and carrier lifetime towards the responsivity of phototransistors. Following this procedure, accurate conclusions can be made as to which combination of materials warrant further research and how to continue improving the performance of this novel class of high-gain, micro-scale photodetectors.

**Experimental Section**

*Device Fabrication*: Monolayer graphene and rubrene single crystals were grown by chemical vapor deposition [34] and physical vapor transport [35] respectively. Full details of material growth and device fabrication are provided in Supplementary Section S1 and S2.

*Photocurrent and PL Measurements:* Phototransistor devices were housed in a vacuum probe station ($10^{-3} mbar$) with a fused silica viewport for photocurrent measurements. A xenon lamp and monochromator with variable low-pass filters (Newport TLS300X) provided spectrally tunable, collimated light incident over the entire sample. We adjusted optical power levels using a series of neutral density filters. A mechanical shutter (Thorlabs SHB1T) modulated light signals and power densities were calibrated using a photodiode (Thorlabs S130CV) before and after each dataset run. Excluding spectral scans, $\lambda = 500nm$ for all measurements. PL spectra were excited in atmospheric conditions using a 532nm laser through a microscope objective (numerical aperture = 0.5). Rubrene-graphene channel dimensions are $L = 5\mu m, W = 91\mu m$, except in Figure 3a.

*Absorbance Calculations:* From Figure 2e, $A = 0.207$ for the rubrene-graphene transistor if $\lambda = 500nm$. The methodology of our absorbance calculations is shown in Figure S4 and S6.

**Supporting Information**
Supporting Information is available from
http://onlinelibrary.wiley.com/doi/10.1002/adma.201702993/full

**Acknowledgements**
S. Russo and M.F. Craciun acknowledge financial support from EPSRC (Grant 464 no. EP/J000396/1, EP/K017160/1, EP/K010050/1, EP/G036101/1, EP/M001024/1, 465 EP/M002438/1), from Royal Society international Exchanges Scheme 2012/R3 and 466 2013/R2 and from European Commission (FP7-ICT-2013-613024-GRASP). The authors would like to thank Paul Wilkins for technical assistance in designing and building the vacuum chamber probe station used for all photocurrent measurements and Dr Dominique J. Wehenkel for assisting with initial photocurrent measurements and for suggesting to measure the length-dependence of responsivity.

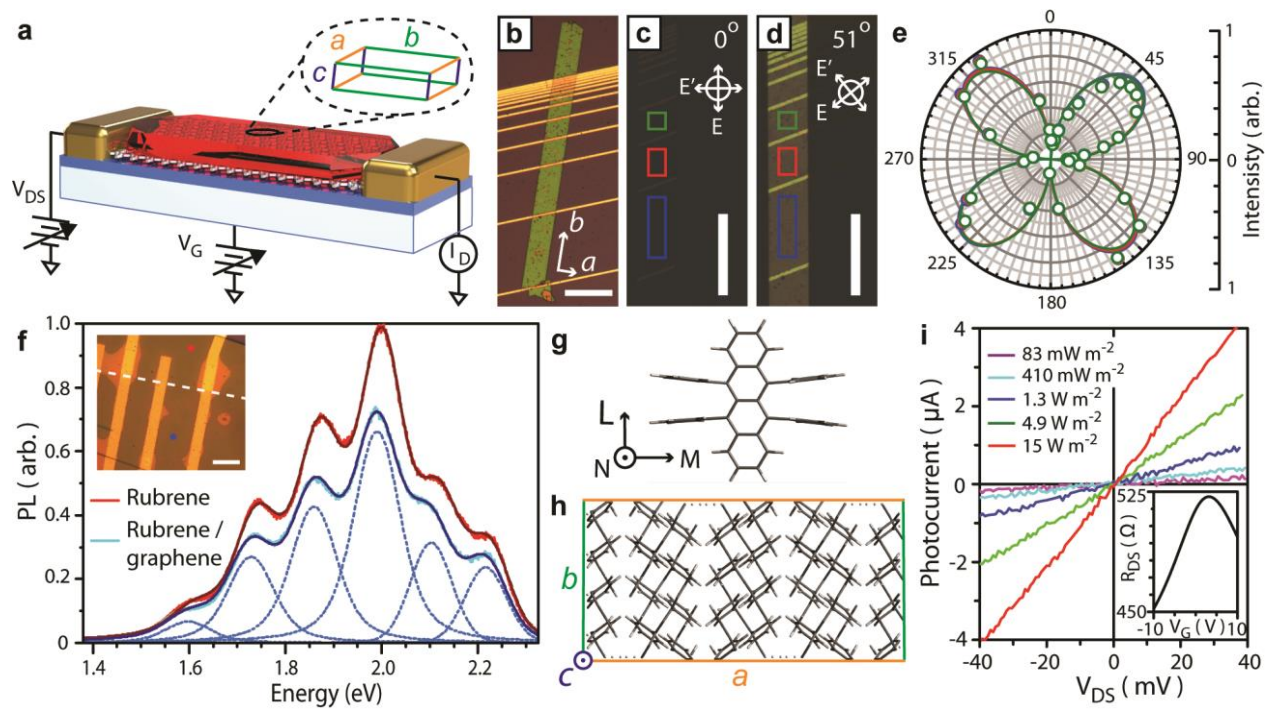

**Figure 1.** Characterization of rubrene-graphene interfaces. a) Schematic of a rubrene-graphene phototransistor on SiO$_2$/Si with a measurement circuit diagram and inset denoting crystallographic axes. b) Un-polarized and c), d) cross-polarized optical micrographs of a rubrene-graphene interface (Scale bars: 200µm). Colored squares denote the regions analyzed in e) polar plots of the greyscale brightness. f) Photoluminescence spectra of a rubrene single crystal at regions with (blue) and without (red) underlying graphene. Dashed peaks are Voigt functions fit to the blue spectra. Inset: Micrograph image showing the location of each PL scan (circles) and boundary between rubrene and rubrene-graphene (dashes). Scale bar: 25µm. g) Molecular and h) crystalline structure of rubrene at room temperature (CSD-QQQCIG08).[26] Axis notation conforms to charge transport and photoluminescence studies.[22, 25] i) Photocurrent as a function of source-drain voltage for various illumination intensities ($V_G = 0V$). Inset: Resistance vs gate voltage sweep of the same channel in dark conditions.

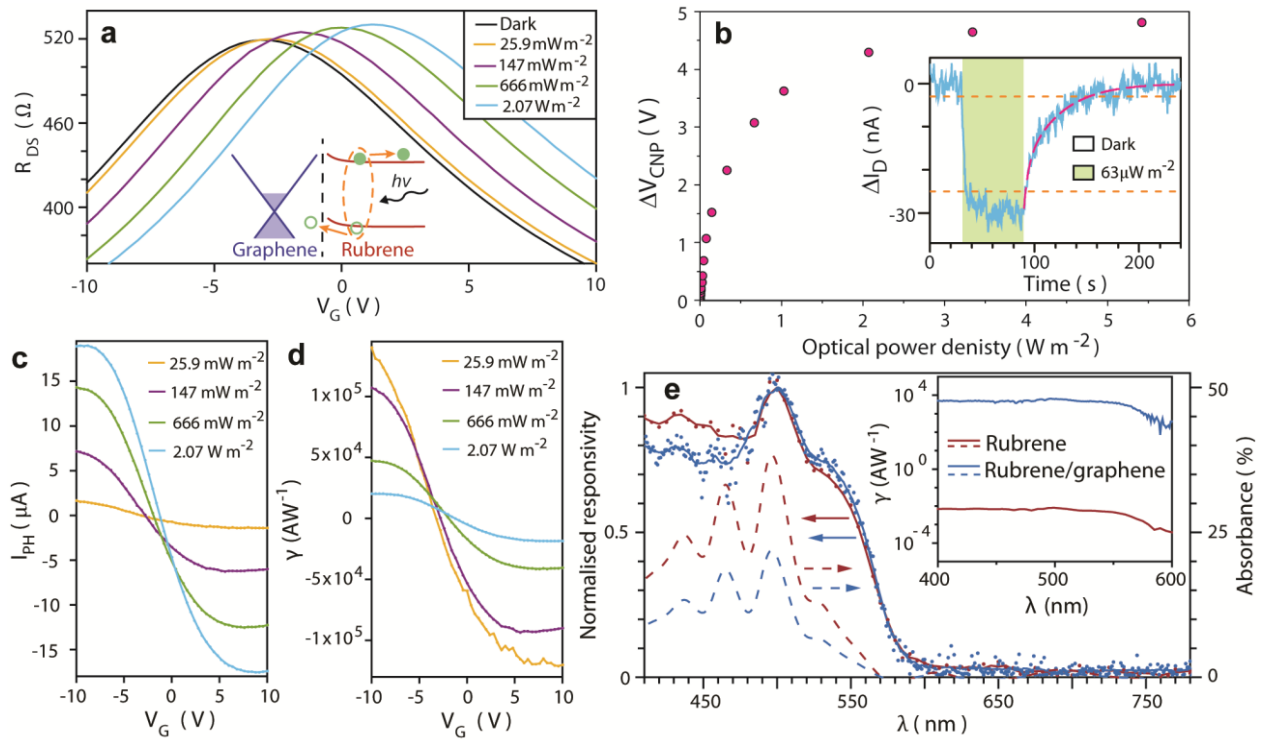

**Figure 2.** Photo-gating in rubrene-graphene interfaces. a) Channel resistance vs gate voltage in dark conditions and under various optical power densities. Inset: Schematic of charge transfer at the rubrene-graphene interface. b) Shift in the charge neutrality point of graphene as a function of optical power density. Inset: 20-run average of the transient response of a 5μm rubrene-graphene channel to 60 seconds of illumination. Dashed lines denote 10% and 90% thresholds of the steady-state shift in current (orange) and a bi-exponential decay fit of the return to dark conditions (pink). c) Photocurrent and d) responsivity as a function of gate voltage. e) Responsivity spectra of rubrene (red) and rubrene-graphene (blue) transistors. Spectra are normalized to the maximum of each dataset (smoothed via adjacent averaging, solid lines). Dashed lines are the simulated net absorbance of each rubrene crystal (see Supplementary Section S3). Inset: Responsivity spectra for rubrene and rubrene-graphene transistors.

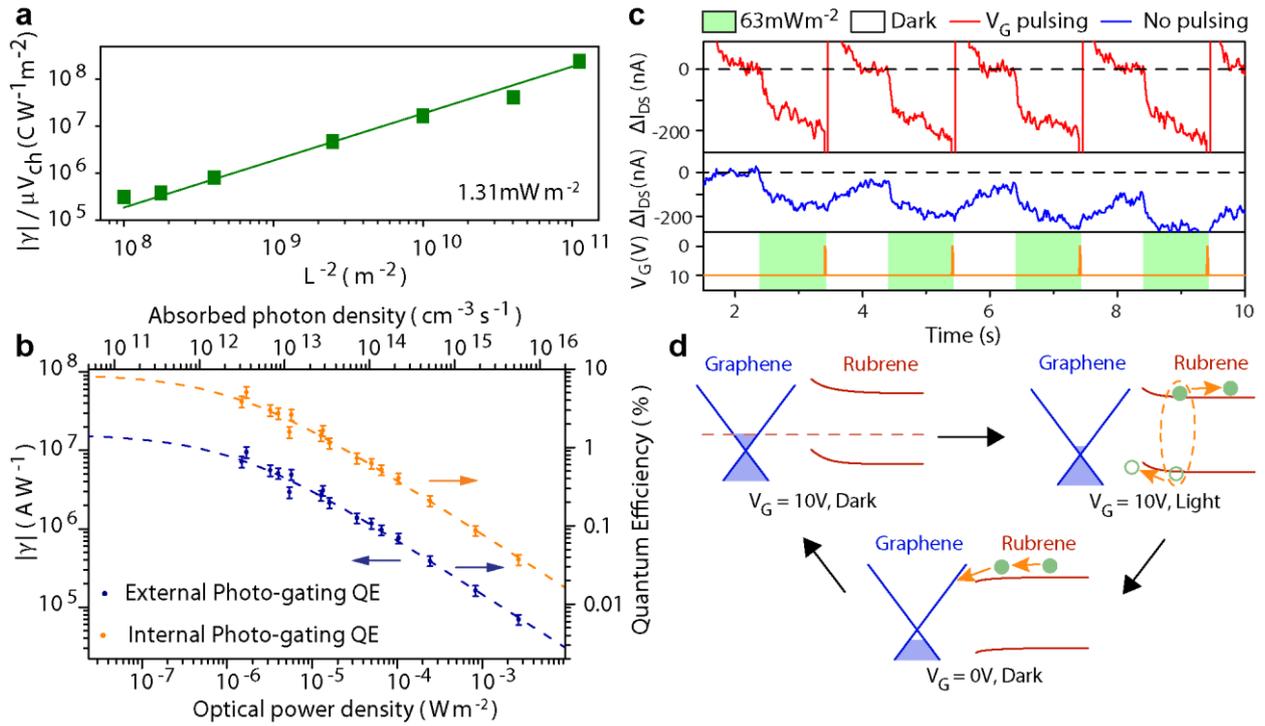

**Figure 3.** Photo-response in rubrene-graphene phototransistors. a) Length scaling of the photo-response in a single rubrene-graphene interface normalized with respect to the charge carrier mobility and potential difference of/across each channel segment (Figure S8 and S9). b) Responsivity vs optical power density and absorbed photon density (blue points, left y-axis) from the averaged response of a 5μm channel to 20 illumination cycles. External (blue points, right y-axis) and internal (orange points, right y-axis) photo-gating quantum efficiencies are calculated from the same dataset. c) Transient photo-response of a rubrene-graphene phototransistor relative to dark current levels (dashes) under light modulated at 0.5Hz with (red) and without (blue) application of gate voltage pulses. Current spikes due to gate pulsing are readily removed with filtering circuitry. d) Schematic band diagrams illustrate the charge transfer dynamics across at each stage of the light modulation cycle.

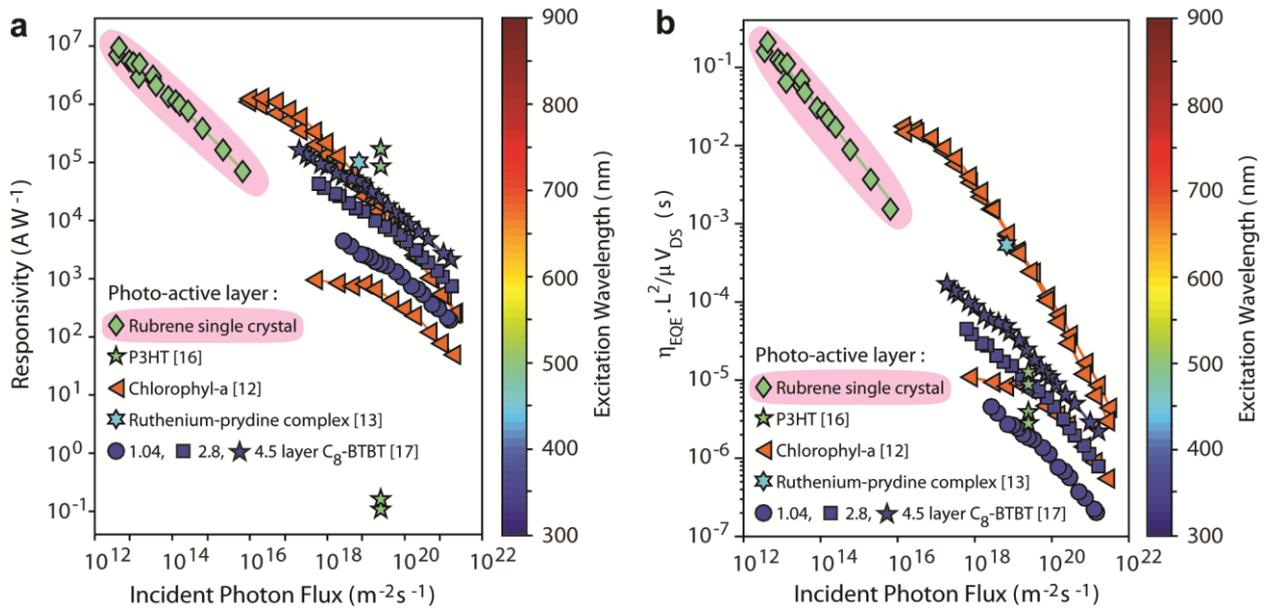

**Figure 4.** Performance metrics of organic semiconductor-graphene phototransistors. Plots catalogue the power dependence of rubrene single crystal-graphene phototransistors with all other relevant studies using a) responsivity and b) the external quantum efficiency normalized to extraneous parameters as figures of merit. Solid lines connect data taken from a single device, marker colors denote the excitation wavelength used and data from this work is highlighted (pink). Table S2 provides detailed citations, Figure S11 shows plots that include inorganic photo-active layers.

**Keywords:** graphene; high quantum efficiency; phototransistor; rubrene single crystal; visible wavelength.

G. F. Jones, R. M. Pinto , A. De Sanctis, V. K. Nagareddy, C. D. Wright, H. Alves, M. F. Craciun , S. Russo*

**Title** Highly efficient rubrene-graphene charge transfer interfaces as phototransistors in the visible regime